\newcommand{\rem}[1]{}
\documentclass[prl,twocolumn,showpacs,preprintnumbers,amsmath,amssymb,floatfix]{revtex4}
\usepackage{hyperref}
\usepackage{epsfig}

\begin{document}

\title{Exactly quantized dynamics of classical incommensurate sliders}

\author{Andrea Vanossi}
\affiliation{CNR-INFM National Research Center S3 and Department
of Physics, \\University of Modena and Reggio Emilia, Via Campi
213/A, 41100 Modena, Italy}
\author{Nicola Manini}
\author{Giorgio Divitini}
\affiliation{Dipartimento di Fisica and CNR-INFM, Universit\`a di Milano,
Via Celoria 16, 20133 Milano, Italy}
\affiliation{International School for Advanced Studies (SISSA), Via Beirut 2-4,
I-34014 Trieste, Italy}
\author{Giuseppe E. Santoro}
\author{Erio Tosatti}
\affiliation{International School for Advanced Studies (SISSA), Via Beirut 2-4,
I-34014 Trieste, Italy}
\affiliation{INFM Democritos National Simulation Center}
\affiliation{International Centre for Theoretical Physics
(ICTP), P.O.Box 586, I-34014 Trieste, Italy}

\date{\today}

\begin{abstract}
We report peculiar velocity quantization phenomena in the classical motion 
of an idealized 1D solid lubricant, consisting of a harmonic chain
interposed between two periodic sliders.
The ratio $v_{cm}/v_{ext}$ of the chain center-of-mass velocity to the
externally imposed relative velocity of the sliders stays pinned to exact
``plateau'' values for wide ranges of parameters, such as sliders
corrugation amplitudes, external velocity, chain stiffness and dissipation,
and is strictly determined by the commensurability ratios alone.
The phenomenon is explained by one slider rigidly dragging 
the kinks that the chain forms with the other slider.
Possible consequences of these results for some real systems are discussed.
\end{abstract}

\pacs{
68.35.Af, 
05.45.Yv, 
62.25.+g, 
62.20.Qp, 
81.40.Pq, 
46.55.+d  
}

\maketitle

\begin{figure}
\epsfig{file=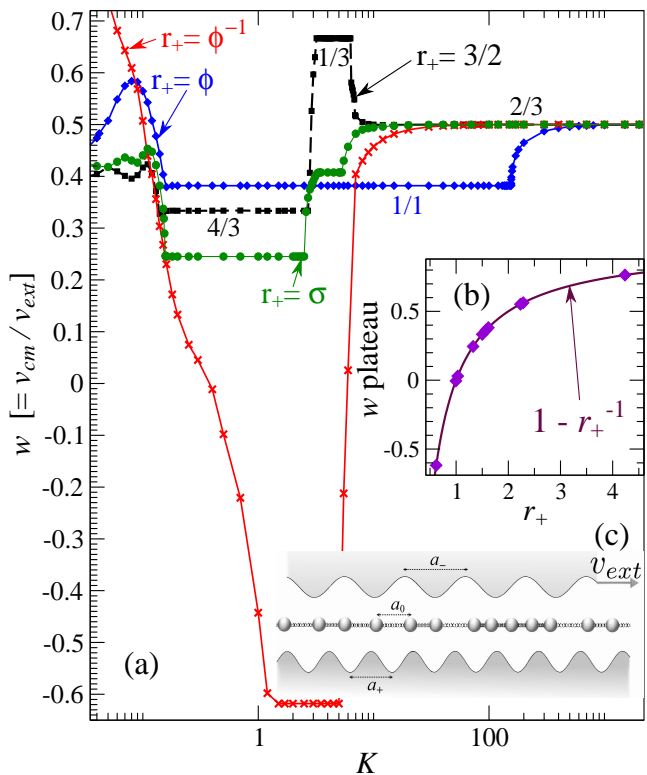,width=8.5cm,angle=0,clip=}
\caption{\label{velcm:fig}
(Color online)
(a) Average drift velocity ratio $w=v_{cm}/v_{ext}$ of the chain as a
function of its spring stiffness $K$ for different length ratios
$(r_+,r_-)$: commensurate $(3/2,9/4)$, golden mean (GM) $(\phi,\phi^2)$,
spiral mean (SM) $(\sigma,\sigma^2)$ ($\sigma \sim 1.3247\dots$ is the
solution of $\sigma^3=\sigma+1$), and $(\phi^{-1},\phi)$.
The plateau labeling is explained in the text.
Here $\gamma=0.1$, $v_{ext}=0.1$, and periodic boundary conditions
\protect\cite{PBC:nota} are used.  The $(\phi,\phi^2)$ 1/1 plateau value is
$w=0.381966\dots$, identical to $1-\phi^{-1}$ to eight decimal places.
(b) The main plateau speed $w$ as a function of $r_+$.
(c) A sketch of the model.
}
\end{figure}

In this Letter we report surprising exact velocity quantization phenomena in a
one-dimensional (1D) non-linear model inspired by the tribological problem of two
sliding surfaces with a thin solid lubricant layer in between \cite{Urbakh,VanossiPRL}.
The model layer consists of a chain of $N$ harmonically
interacting particles interposed between two rigid generally
(but not necessarily) incommensurate sinusoidal substrates (the two ``sliding crystals'',
sketched in Fig.~\ref{velcm:fig}c) externally driven
at a constant relative velocity $v_{ext}$.
The equation of motion of the $i$-th particle is:
%
%
\begin{eqnarray} \label{eqmotion:eqn}
m\ddot{x}_i &=&  -\frac{1}{2} \left[ F_+ \sin{k_+ (x_i-v_+t)} + F_- \sin{k_-(x_i-v_-t)}\right] \nonumber \\
&+& K (x_{i+1}+x_{i-1}-2x_i) - \gamma \sum_{\pm} (\dot{x}_i - v_{\pm}) \;,
\end{eqnarray}
where $m$ is the mass of the $N$ particles, $K$ is the chain spring constant, and
$k_{\pm}=(2\pi)/a_{\pm}$ are the wave-vector periodicities of potentials representing
the two sliders, moving at velocities $v_{\pm}$.
We set, in full generality, $v_+$ = 0 and $v_{ext} = v_- - v_+$.
$\gamma$ is a phenomenological parameter substituting for various sources
of dissipation, required to achieve a stationary state, but otherwise
with no major role in the following.
$F_{\pm}$ are the amplitudes of the forces due to the sinusoidal
corrugation of the two sliders (we will commonly use $F_-/F_+=1$ but we
checked that our results are more general).
We take $a_+=1$, $m=1$, and $F_+=1$ as our basic units. 
The relevant length-ratios \cite{vanErp99,Vanossi00} are therefore $r_{\pm}=a_{\pm}/a_0$; 
we will take, without loss of generality, $r_->r_+$, and confine our attention mostly to cases
with $r_+>1$.
%
%
The inter-particle equilibrium length $a_0$, not entering explicitly the
equation of motion (\ref{eqmotion:eqn}), appears only via the boundary
conditions, which are taken to be periodic (PBC) \cite{PBC:nota},
$x_{N+1}=x_1+N \, a_0$, 
to enforce a fixed density condition for the chain \cite{BraunBook}, with a
{\em coverage} $r_+$ of chain atoms on the denser substrate.
Previous studies on a related model \cite{VanossiPRL} achieved the sliding
through the application of a constant driving to one of the two substrates,
via an additional spring. That procedure obscures the surprising
quantization phenomena which are instead uncovered when sliding occurs
with a constant velocity $v_{ext}$.
 
Upon sliding the substrates, $v_{ext} \neq 0$, the lubricant chain slides too.
However, it generally does so in a very unexpected manner: 
the time-averaged chain velocity $w=v_{cm}/v_{ext}$,
is generally {\it asymmetric}, namely different from 1/2.
Even more surprisingly, $w$ is exactly {\it quantized}, for large parameter
intervals, to plateau values that depend solely on the chosen
commensurability ratios.
The asymmetrical $w$-plateaus are generally very stable, and insensitive to
many details of the model, due to their intrinsically topological nature; we
show that they are the manifestation of a certain density of solitons in
the lubricant which are set into motion by the external driving.

\rem{

Equation~(\ref{eqmotion:eqn}) describes a generalized Frenkel-Kontorova
(FK) model \cite{BraunBook}.
In the standard FK model (chain plus a single periodic substrate potential),
the static response to an external force attempting to slide
the chain is known to depend strongly on the value of commensurability $r$.
Rationals $r$ always lead to pinning: a finite force is necessary to set
the chain into sliding motion.
Irrationals $r$ display an Aubry transition
between a pinned phase for soft chains $K < K_{\rm Aubry}$,
and an unpinned regime for hard chains $K > K_{\rm Aubry}$ \cite{BraunBook,Aubry83}.
The critical stiffness strongly depends on the value and on the
number-theory properties of the irrational $r$; in particular, $K_{\rm
Aubry}$ is smallest for $r=\phi\equiv(1+\sqrt{5})/2$, the Golden Mean (GM)
\cite{Aubry83}.
In a sense, the GM represents the ``most irrational'' incommensuration,
with the largest unpinned phase domain as $K$ decreases from $\infty$ to
$K_{\rm Aubry}$.
This finding goes hand-in-hand with the well-known number-theoretical
property that, among irrationals, the GM is the hardest to approximate by
rationals, a property ultimately connected to its continued-fraction
expansion coefficients being all equal to unity \cite{Gutzwiller90}.

Explicit solution of the dynamics of our richer 1D solid-lubricant-solid FK model, presented below,
shows that the addition of a further length scale completely changes the picture,
turning the GM behavior into the {\em closest} to commensurate.

}     	

We now turn to illustrating these results.
We integrated the equations of motion (\ref{eqmotion:eqn}) starting from
fully relaxed springs ($x_i=i\, a_0$, $\dot{x}_i=v_{ext}/2$), by a standard
fourth-order Runge-Kutta method.
After an initial transient, the system reaches its dynamical stationary
state, at least so long as $\gamma$ is not exactly zero.
Figure~\ref{velcm:fig}a shows the resulting time-averaged center-of-mass (CM) velocity
$v_{cm}$ as a function of the chain stiffness $K$ for four representative
$(r_+,r_-)$ values.
%
We find that $w$ is generally a complicated function of $K$, with flat
plateaus and regimes of continuous evolution, in a way which is
qualitatively similar for different cases.
The main surprise is that all plateaus show perfectly flat
$w=v_{cm}/v_{ext}$ that are constant (quantized) to all figures of numerical
accuracy, the precise value strikingly independent not only of $K$,
but also of $\gamma$, $v_{ext}$, and even of $F_-/F_+$.
Open-boundary simulations show moreover that the PBC used are not crucial
to the plateau quantization, which occurs even for a lubricant of finite
size and not particularly large $N$, such as a hydrocarbon chain molecule would be.
%

\begin{figure}
\centerline{
\epsfig{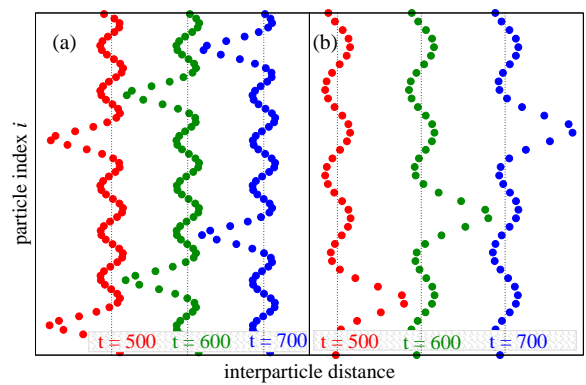}
}
\caption{\label{kinks:fig}
(Color online)
Snapshots of the distance between neighbor lubricant particles in the
chain $x_i-x_{i-1}$ at three successive time frames.
All parameters as in Fig.~\ref{velcm:fig}, except $r_-\simeq 10.36$, and
$K=10$ (inside the main plateau).
(a) $r_+=1.031$ (kink density $\delta/a_+= 0.031$); (b) $r_+=0.995$
(anti-kink density $|\delta|/a_+ = 0.005$).
}
\end{figure}

To explore the origin of the universality of $w$, we analyze the dynamics
for a large number of values of $(r_+,r_-)$, and observe that:
(i) at least one velocity plateau as a function of $K$ occurs 
for any $(r_+,r_-)$; 
(ii) additional narrower secondary plateaus often arise for stiffer
lubricant (larger $K$, see Fig.~\ref{velcm:fig});
(iii) the velocity ratio $w$ of the first plateau found for increasing
$K$, satisfies $w=1-r_+^{-1}$ for a large range of $(r_+,r_-)$.

We can understand these results as follows. 
Consider initially the situation of quasi-commensuration of the chain to 
the $a_+$ substrate: $r_+=1+\delta$, with small $\delta$. 
This induces a density $\rho_{sol}=\delta/a_+$ of solitons (or
kinks, essentially substrate minima holding two particles, rather than one)
\cite{BraunBook}.
The second, less oscillating $a_-$ slider, which moves at velocity
$v_{ext}$, will {\em drag} the kinks along: $v_{sol}=v_{ext}$.
If $\rho_0 = 1/a_0 = r_+/a_+$ is the linear density of lubricant particles,  
mass transport will obey $v_{cm} \rho_0 = v_{sol} \rho_{sol}$. 
This yields precisely
$w=v_{cm}/v_{ext}=\rho_{sol}/\rho_0=\delta/r_+=1-r_+^{-1}$.  Thus the
exact plateaus arise because the smoother slider (whose exact
periodicity $r_->r_+$ is irrelevant) drags the kinks, of given density, at
its own full speed $v_{ext}$, as illustrated Fig.~\ref{kinks:fig}a.
As shown in Fig.~\ref{velcm:fig}b, this physics extends even to large
$|\delta|\sim 1$, where no individual kink can be singled out.
This works even for $\delta<0$ (anti-kinks), where, remarkably, the
lubricant CM moves in the {\em opposite} direction with respect to the
driving $v_{ext}$ ($w<0$, see Fig.~\ref{velcm:fig}a for $r_+=\phi^{-1}$):
exactly as holes in a semiconductor, anti-kinks (carrying a negative
``charge'') moving at velocity $+v_{ext}$ effectively produce a backward
net lubricant motion.
The motion of the anti-kinks (regions of increased inter-particle
separation) is illustrated in Fig.~\ref{kinks:fig}b.

\begin{figure}
\epsfig{file=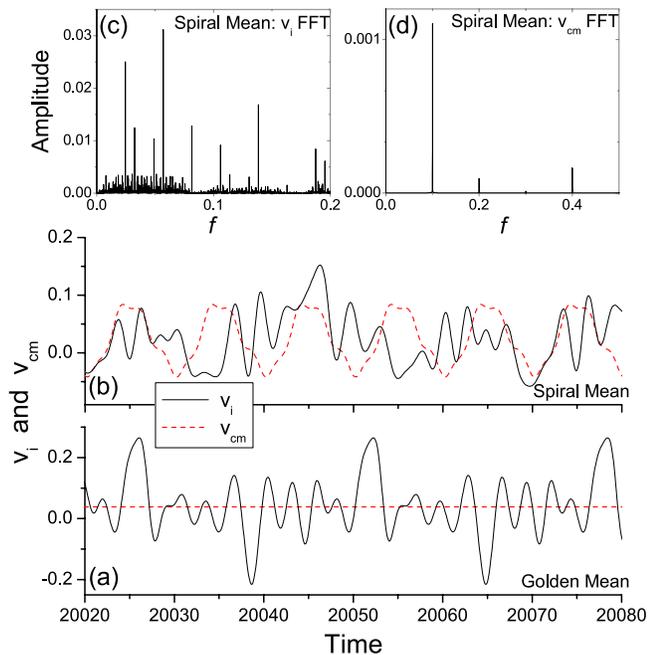,width=8.5cm,angle=0,clip=}
\caption{\label{evol:fig}
(Color online) Time evolution of a particle velocity $\dot{x}_i$, and of the chain CM velocity
$v_{cm}$ (fluctuations rescaled by a factor 50), for the GM (a) and SM (b) cases of Fig.~\ref{velcm:fig}.
Amplitudes of the Fourier spectrum of $\dot{x}_i(t)$ (c) and of $v_{cm}(t)$ (d) for $r=\sigma$.
Individual particle spectra have identical amplitudes, and
differ only in the phases, which leads to a remarkable cancellation in the
$v_{cm}$ power spectrum. Here $K=1$, $\gamma=0.1$, and $v_{ext}=0.1$.
}
\end{figure}

The motion of particles is instructive.
Figure~\ref{evol:fig} plots the time evolution of the velocity of a single
chain particle $\dot{x}_i$, and of $v_{cm}$, for a value of $K$ inside a
plateau, $K=1$, for $r_+=\phi$ and $r_+=\sigma$ of Fig.~\ref{velcm:fig}.
Here two clear kinds of behavior emerge.
Remarkably, the single-particle motion in the GM plateau is perfectly
time-periodic.
A similar periodic dynamics is found, in appropriate regimes, for all
rational and {\em quadratic} irrational $(r_+,r_-)$ values we tested.
The SM motion is definitely not periodic.
The Fourier spectrum of the particle motion, shown in
Fig.~\ref{evol:fig}c, confirm that the SM yields only quasi-periodic
orbits with two prominent incommensurate frequencies $f_+$ and $f_-$.
However, a phase-cancellation between the Fourier spectra of different
chain particles (all having the same amplitude spectrum, with different
phases) yields strictly periodic CM motion. 
Periodic CM oscillations around an exactly quantized drift velocity is a
common feature of all plateaus in the chain dynamics.
These periodic oscillations can be understood as the solitons moving at
velocity $v_{ext}$ encountering a periodic Peierls-Nabarro potential
\cite{BraunBook} of period $a_+$.
The corresponding frequency of encounter $v_{ext}/a_+$ is clearly visible
in Fig.~\ref{evol:fig}d.
We observe that the Peierls-Nabarro barrier vanishes (strictly constant
$v_{cm}$) for all cases where the particle motion is periodic.

To understand why single particles may move periodically in some plateaus,
let $f_+= v_{cm}/a_+$ and $f_-= (v_{ext}-v_{cm})/a_-$ be the average
encounter frequency of a generic particle with the two substrate periodic
corrugations.
Whenever these two frequencies are mutually commensurate,
\begin{equation} \label{matching:eqn}
n_+ f_+=n_-f_- \qquad({\rm integers} \ n_{\pm})
\,,
\end{equation}
each particle is driven by the two generally incommensurate corrugations in
an effectively periodic way, and undergoes a periodic oscillation, of period
$T=n_-/f_+= n_+/f_-$.
This is observed in the Fig.~\ref{velcm:fig} examples: $r_+=\phi$ yields
$n_+/n_-=1/1$ and $r_+=3/2$ yields $n_+/n_-=4/3$.  The
plateaus in Fig.~\ref{velcm:fig} are labeled accordingly.
When $f_+$ and $f_-$ are incommensurate, the motion is quasi-periodic, as
illustrated by the SM case of Fig.~\ref{evol:fig}.
By definition, $f_+/f_-=(r_-/r_+)\,w/(1\!-\!w)$: for any given plateau
velocity $w$, suitable choices of $r_-/r_+$ can make the two frequencies
either commensurate, as in Eq.~\eqref{matching:eqn}, or incommensurate.
In particular, for the main plateau we have $f_+/f_-= r_- (r_+\!-\!1)/r_+$,
which can be made rational for any choice of $r_+$, by choosing a suitable
$r_-=r_+(r_+\!-\!1)^{-1}\, n_-/n_+$.
For example, we verified that, for $r_+=\sigma$ and $r_-=\sigma
(\sigma\!-\!1)^{-1}$, individual particles do oscillate periodically in the
main plateau.
 
\rem{
In particular, for $r_-=r_+^2$ \cite{VanossiPRL,vanErp99,Vanossi00}, the
relations sketched above imply that in the main plateau the orbits are
periodic for any $r_+=1/2 +(1/4+n_+/n_-)^{1/2}$, while are quasi-periodic
for any other $r_+$.
Other minor plateaus, often arising for $K\gtrsim 3$ as illustrated in
Fig.~\ref{velcm:fig}, are associated to either periodic oscillations or
quasi-periodic motion.
}

Low driving velocities $v_{ext}$ are beneficial to the appearance and width
of plateaus.
For increasing $v_{ext}$, the plateaus shrink and eventually disappear,
still remaining exact while they do so.
The critical $v_{ext}$ where the plateaus end depends on $K$, but is
usually smaller than unity, for the parameters of Fig.~\ref{velcm:fig}.
The $1/1$ plateau of the GM case 
is especially wide and robust against an increase of $v_{ext}$ (for the parameters of
Fig.~\ref{velcm:fig} and $K=4$, up to $v_{ext}\simeq 1.5$) and
other perturbations.
This seems related to $\phi$ being uniquely associated to the equal
drive-frequency ratio, $n_+/n_-=1/1$:
$\phi$ appears therefore, in the present dynamical context, as the ``most
{\em commensurate}'' irrational, at variance with static pinning in the
standard FK model, where the opposite is true \cite{BraunBook,Aubry83}.
%
%
%

\begin{figure}
\epsfig{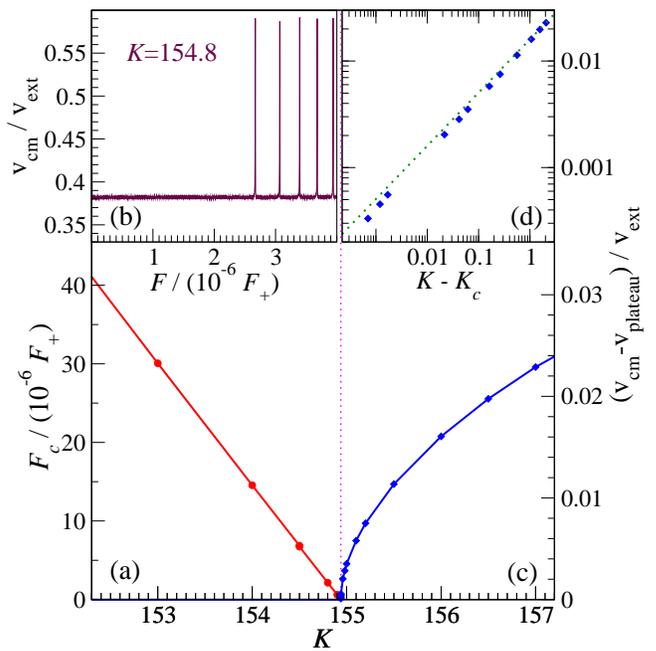}
\caption{\label{critical:fig}
(Color online)
Dynamical depinning force $F_c$ for the GM plateau (a), extracted by a slow
adiabatic increase of the force $F$ applied to all particles, until
intermittencies appear, signaling collective slips (b).  The critical
behavior of the average velocity for $K\to K_c$ (c,d), showing the typical
square-root singularity (dotted line) associated to type-I intermittencies.
Here $\gamma=0.1$, and $v_{ext}=0.1$ were used.
}
\end{figure}

The finding of exact plateaus implies a kind of ``dynamical
incompressibility'', namely identically null response to perturbations or
fluctuations trying to deflect the CM velocity away from its quantized
value.
In order to probe the robustness of the plateau attractors, we introduced
an additional constant force $F$ acting on all particles in the chain,
trying to alter the force-free sliding in the plateau regime.
As expected, as long as $F$ remains sufficiently small, it does
perturb the single-particle motions but has no effect whatsoever on $w$,
which remains exactly pinned to the attractor ($v_{cm}\equiv v_{plateau}$).
The plateau dynamics is only abandoned above a critical force $F_c$.
This dynamical depinning takes place through a series of type-I
intermittencies \cite{Berge84}, as shown in Fig.~\ref{critical:fig}b
where $v_{cm}(t)$ is plotted against a slow adiabatic ramping of $F$.

A precise value of $F_c$ can be obtained by ramping $F$ with time with a
gentle enough rate of increase or, alternatively, by a Floquet-Lyapunov
linear stability analysis \cite{Jose98}, to be shown elsewhere.
The value of $F_c$ is a function of the parameters, and
$F_c$ vanishes linearly when $K$ approaches the border $K_c$ of the plateau,
as in Fig.~\ref{critical:fig}a.
The depinning transition line $F_c$, ending at $K=K_c$, is a `first-order'
line, with a jump $\Delta v$ in the average $v_{cm}$ and a clear hysteretic
behavior (not shown) as $F$ crosses $F_c$. As can be expected,
$\Delta v$ decreases to $0$ as $K$ increases towards $K_c$.
Thus $K=K_c$ represents a genuine non-equilibrium critical point, where
the sliding chain enters or leaves a dynamical orbit.
The precise value of $K_c$ depends on parameters such as $v_{ext}$ and $\gamma$;
however, its properties do not.
As $K$ approaches $K_c$ from above (no external force), $v_{cm}$ approaches
$v_{plateau}$ in a critical manner, as suggested in Fig.~\ref{velcm:fig}.
This is detailed in Fig.~\ref{critical:fig}(c,d), where the critical behavior
is shown to be $\Delta v\propto (K-K_c)^{1/2}$, the value typical of
intermittencies of type I \cite{Berge84}.
For $K\gtrsim K_c$, in fact the chain spends most of its time moving 
at $v_{cm}(t) \simeq v_{plateau}$, except for short bursts at regular time-intervals
$\tau$, where the system as a whole jumps ahead by $a_0$, i.e.\ an extra
chain lattice spacing (collective slip).
The characteristic time $\tau$ between successive collective slips
diverges as $\tau\propto(K-K_c)^{-1/2}$ for $K\to K_c$, consistent with
the critical behavior of $w$.
%
%
We verified that the $w$-plateaus for more general values of $r_+$ and of
$r_-$ show the same kind of infinite stiffness, and a critical decrease of
$F_c$ near the plateau edge, similar to that of Fig.~\ref{critical:fig}a
for the GM.

\rem{
Over substantial intervals of the dynamical parameters and of $r'$, the
value of the velocity ratio $w$ is seen to depend only on $r$.
The nature (either periodic or quasi-periodic) of the lubricant
oscillations does not affect $w$.
%
The rigid velocity quantization seems connected to a mechanism of discrete
particle locking/unlocking between the maxima of the slider potential, but
the exact origin of the universal plateau phenomenon remains mysterious.
}

The phenomena just described for a model 1D system are quite extraordinary; 
it would be fascinating if they could be observed in real systems.
Nested carbon nanotubes \cite{Zhang}, or confined one-dimensional nanomechanical systems 
\cite{Toudic_06}, are one possible arena for the phenomena described in this Letter.
Though speculative at this stage, one obvious question is what aspects of
the phenomenology just described might survive in two-dimensions (2D),
where tribological realizations, such as the sliding of two hard
crystalline faces with, e.g., an interposed graphite flake, are
conceivable.
Our results suggests that the lattice of discommensurations -- a Moir\'e
pattern-- formed by the flake on a substrate, could be dragged by the other
sliding crystal face, in such a manner that the speed of the flake as a
whole would be smaller, and quantized. This would amount to the slider
``ironing'' the solitons onward.
Dienwiebel {\it et al.}\ \cite{Dienwiebel04} demonstrated how
incommensurability may lead to virtually friction-free sliding in such a
case, but no measure was obtained for the flake relative sliding velocity.
Real substrates are, unlike our model, not rigid, subject to thermal expansion, etc. 
Nevertheless the ubiquity of plateaus shown in Fig.~\ref{velcm:fig}, 
and their topological origin, suggests that these effects would not remove the phenomenon.
A real-life situation with a distribution of differently oriented crystalline micro-grains,
each possessing a different incommensurability,
is also potentially interesting; each grain, we expect, will tend to
stabilize a certain average CM velocity depending on its incommensurability. 
%
Other realizations or applications inspired by the physics described by our
model might be accessible, notably in grain boundary motion, in the sliding
of optical lattices \cite{optical} or of charge-density-wave systems \cite{Gruener_cdw}.

Acknowledgments -- We are grateful to O.M. Braun for invaluable discussions.  
This research was partially supported by PRRIITT (Regione
Emilia Romagna), Net-Lab ``Surfaces \& Coatings for Advanced Mechanics and
Nanomechanics'' (SUP\&RMAN) and by MIUR Cofin 2004023199, FIRB RBAU017S8R,
and RBAU01LX5H.


\begin{thebibliography}{10}
\bibitem{PBC:nota} 
%
PBC are realized either by (i) approximating $a_-/a_+$ and $r_+=a_+/a_0$
with suitable rational numbers
$N_+/N_-$ and $N/N_+$, or (ii) by using a finite $N$ and
machine-precision values of $r_+$ and $r_-$.
%
Both methods introduce a finite-size error, which can be made arbitrarily
small by choosing suitably large integers $N,N_{\pm}$, and lead to the same
results for large $N,N_{\pm}$, where full incommensurability is restored.
\end{thebibliography}

\end{document}